\documentclass[journal]{IEEEtran}
\IEEEoverridecommandlockouts

\usepackage{hyperref}
\hypersetup{colorlinks,allcolors=black}
\usepackage{amsmath,amssymb,amsfonts}
\usepackage{algorithmic}
\usepackage{graphicx}
\usepackage{textcomp}
\usepackage{xcolor}
\usepackage{comment}
\usepackage{authblk}

\usepackage{gensymb}
\usepackage{color}
\usepackage{graphicx}  
\usepackage{soul}
\usepackage{physics}

\usepackage[ruled,vlined,linesnumbered]{algorithm2e}
\usepackage{comment}
\usepackage[numbers,sort&compress,square]{natbib}
\usepackage{enumerate}
\usepackage[normalem]{ulem}

\usepackage{pdfpages}

\usepackage[margin=.65in]{geometry}

\usepackage[normalem]{ulem}
\usepackage{float}
\usepackage{scalerel}
\usepackage{tikz}
\usetikzlibrary{svg.path}

\usepackage{subcaption}

\usepackage{url}

\def\BibTeX{{\rm B\kern-.05em{\sc i\kern-.025em b}\kern-.08em
    T\kern-.1667em\lower.7ex\hbox{E}\kern-.125emX}}
 
\usepackage[font=small,skip=0pt,labelsep=period,justification=justified]{caption}
    
\usepackage[normalem]{ulem}

\begin{document}

\title{A Literature Review on Mobile Charging Station Technology for Electric Vehicles}
\author{Shahab Afshar, Pablo Macedo, Farog Mohamed, and Vahid Disfani\\
\normalsize{ConnectSmart Research Laboratory, University of Tennessee at Chattanooga, TN 37403, USA} \\
Emails: 
shahab-afshar@mocs.utc.edu, vahid-disfani@utc.edu}

\maketitle
\begin{abstract}
While Electric vehicles (EVs) adoption is accelerating in an unprecedented way, lacking EV charging infrastructure hinders the development of the  EV market. To compensate for these shortcomings, Mobile Charging Stations (MCS) could play a prominent role to accelerate EV penetration by providing charging services with no restrictions on the location and time of the charging process. This paper disseminates information on other papers and technical reports on MCS in the literature. It also discusses the benefits of MCS, its challenges, and finally introduces the research gaps in this area.

\end{abstract}
\begin{IEEEkeywords}
EV charging infrastructure, fast charging, mobile charging station,  off-grid charging, technical benefits.  
\end{IEEEkeywords}

\let\thefootnote\relax\footnotetext{© 2020 IEEE. Personal use of this material is permitted. Permission from IEEE must be obtained for all other uses, in any current or future media, including reprinting/republishing this material for advertising or promotional purposes, creating new collective works, for resale or redistribution to servers or lists, or reuse of any copyrighted component of this work in other works.}

\let\thefootnote\relax\footnotetext{}
\let\thefootnote\relax\footnotetext{This paper has been accepted for presentation at the 2020 IEEE Transportation Electrification Conference \& Expo, to be held in Chicago, Illinois USA from June 24-26, 2020.} 
\section{Introduction}
EV penetration is expanding at a rapid pace. 
More than two million EVs were sold in 2018 and the projection for 2025 is 10 million EVs to be sold. By 2040, it is expected that 57\% of all passenger vehicle sales and over 30\% of the global passenger vehicle fleet be electric \cite{finance2019electric}.
The world’s energy demand for EV could also grow from 20 billion kWh in 2020 to 280 billion kWh in 2030 \cite{engel2018charging}. 
Since the driving range limit is one of the key factors restricting EV penetration, building an adequate number of charging stations to cover the charging demand of all these EVs will be a big concern in the near future. An effective way to keep up with the increasing EV penetration is to use different methods of charging with different features to compensate for each method's shortcomings.

\begin{figure}[htbp]

\centering
\includegraphics[width=1 \linewidth]{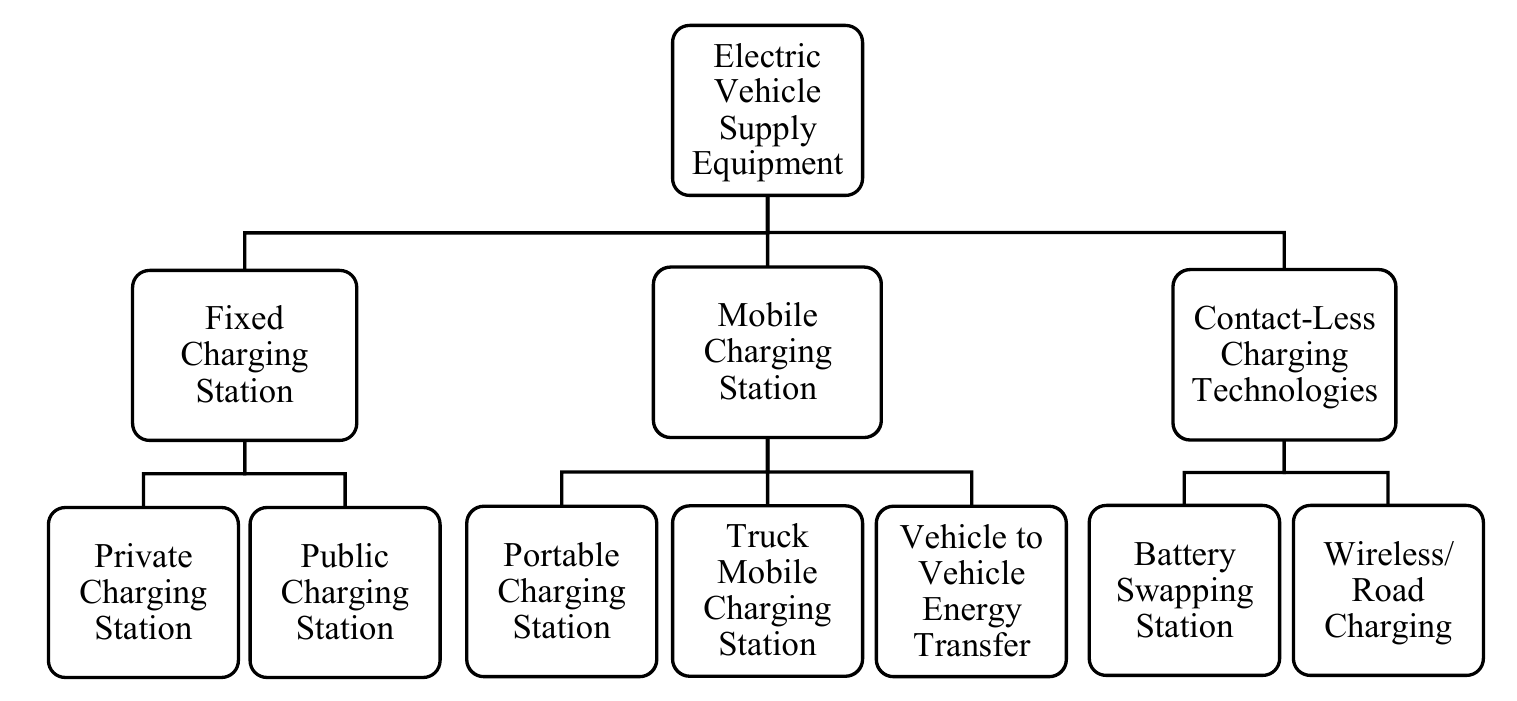}
\caption{Classification of charging methods.}
\label{fig:EVSE Classification}
\end{figure}

This paper presents a literature review on different technologies to supply the charging demand of EVs, including fixed  and mobile charging stations, battery swapping, and wireless charging as classified in Fig. \ref{fig:EVSE Classification}. Due to their popularity, the majority of the existing research works in the literature are focused on fixed charging stations (FCS). Battery swapping and wireless charging lanes which are seldom employed due to their immature technology, relatively large construction costs, and difficulty in standardization \cite{cui2018locating} are also reviewed. The main focus of this review paper is on technology, benefits, and application of mobile charging stations (MCS).

\section{Conventional Charging Technologies}
{To show the logic behind the classification of charging methods and the possibility to merge different charging methods, in this section, we introduce conventional charging technologies.} 
\subsection{Fixed Charging Stations}
\label{sec:FCS}
Fixed charging stations (FCS) are fixed facilities in the shape of a regular electricity plug or a building equipped with one or several charging piles. The power is gained from the electricity grid or local energy generator \cite{atmaja2015electric}. The number of FCS continued to rise in 2018 to an estimated 5.2 million worldwide for light-duty vehicles \cite{outlook2019electric}. Based on their accessibility, FCSs are categorized into private charging stations (PrCS) and public charging stations (PuCS).

Most FCSs are slow chargers at homes and workplaces, often referred to as private charging stations (PrCS). PrCS are installed in locations requiring proprietary access including private residential and business parking facilities accessible only by authorized residents, employees, or patrons \cite{CLINTON2019102255}.

While EV owners can charge at home at low charging rates, the shortage of private parking lots in big cities and the long EV charging time are two important factors which derive the need for public charging stations (PuCSs) for charging of EVs \cite{chauhan2018scheduling}. As a result, in 2018, almost 540,000 publicly accessible chargers (including 150,000 fast chargers) were installed worldwide. With the 156,000 fast chargers for buses, by the end of 2018, there were about 300,000 fast chargers installed globally \cite{outlook2019electric}.

\subsection{Contact-less Charging Technologies}
\label{Sec:contactless}
In addition to conventional charging methods which rely on EVs being connected to charging stations, there are some other technologies such as battery swapping and wireless charging which do not need direct electrical connection between EVs and chargers. 
Battery swapping is a solution that enables EV to be fully recharged in a few minutes instead of hours \cite{noauthor_NIO2_2019}. The technology consists of a device that swaps the depleted battery with a fully charged one. 

Wireless power transfer (WPT) is a technology which allows EVs to be charged on the road through a wireless magnetic connection between EV and the coils implemented on the road. WPT has been recently subject of several studies due to its convenient and safety. Among the studies there are mostly two general approaches to conceive WPT: the inductive power transfer and the capacitive power transfer.

\section{Mobile Charging Station: State-Of-The-Art}
\label{sec:MCS_Research}

A Mobile charging station is a new type of electric vehicle charging equipment which can offer EV charging services at any location or time requested \cite{cui2018mobile}.
MCSs are dispatched in response to two kinds of requests, (i) from overloaded FCS or (ii) from EVs \cite{atmaja2015electric}. Due to the novelty in MCS technology and the attention it receives in the literature recently, this paper focuses on different aspects this technology.

\subsection{Types and technologies}

\begin{figure*}[ht]
  \centering
\begin{subfigure}{.4\textwidth}
  \centering
  \includegraphics[width=.9\linewidth]{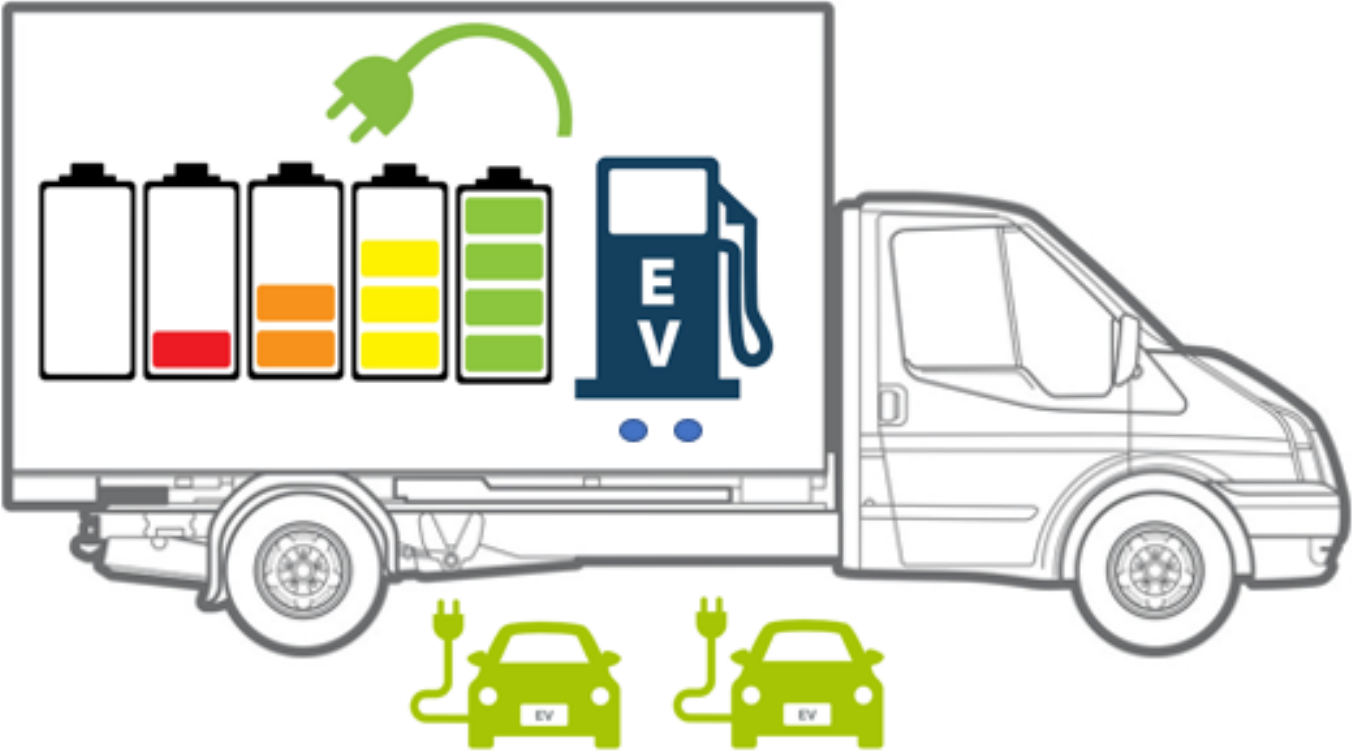}  
  \caption{}
  \label{fig:sub-first}
\end{subfigure}
\begin{subfigure}{.4\textwidth}
  \centering
  \includegraphics[width=.9\linewidth]{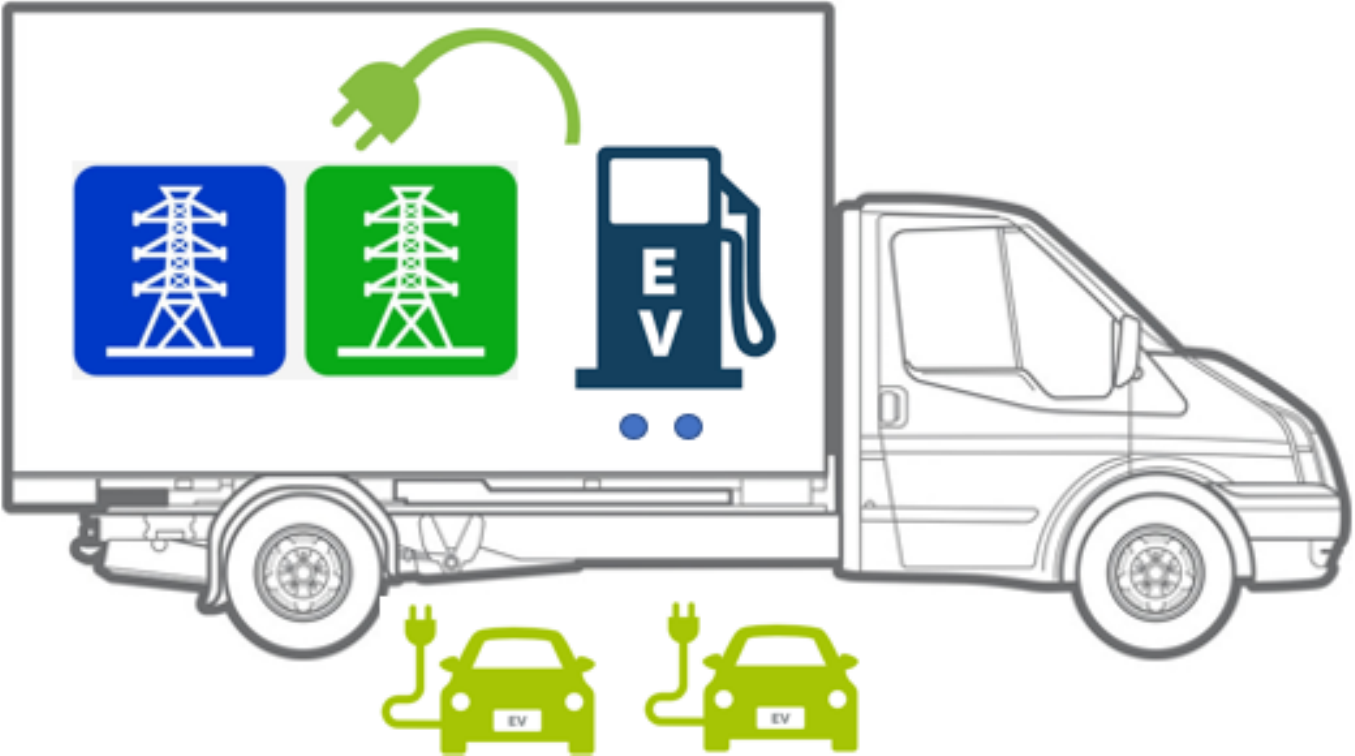}  
  \caption{}
  \label{fig:sub-second}
\end{subfigure}
\caption{Truck mobile charging stations: (a) TMCS with BESS, (b) TMCS without BESS}
\label{fig:TMCS}

\end{figure*}

There are different configurations for MCS. Truck Mobile Charging Stations (TMCSs) are electric or hybrid vehicles, \textit{e.g.} a truck or a van, equipped with one or more charging piles, which can travel a distance in a certain range to charge EVs. There are two types of TMCS as shown in Fig. \ref{fig:TMCS}. Some TMCSs are not equipped with any types of energy storage and just visit FCSs which need more points of connection and connect to their plugs to provide more points of connection for additional EVs \cite{yang2013mobile}. {In other words, the TMCSs's power sources are coming from the connection to the power network via FCSs inlet.} Other TMCSs are equipped with some mounted battery energy storage systems (BESS) \cite{atmaja2015energy}. These TMCSs could visit EVs at any location (e.g. where they are parked) to charge them but need to charge their BESS at a charging station for their next missions. {Portable Mobile Charging Stations (PMCSs) include a mobile BESS which is towed or carried by a vehicle–as opposed to the BESS mounted on TMCS, which allows for its stand-alone operation of PMCS.}

\subsection{Benefits of MCS}

MCSs address several shortcomings of FCSs to improve EV driving experience. Furthermore, reducing negative effects of FCSs on the power grid and motivating investors to invest in FCSs are other beneficial effects of MCSs.

\subsubsection{Benefits to EV owners}
Range anxiety, charging time, and charging availability are essential parameters affecting EV driving experience. 
\begin{enumerate}
\item[i.] Range Anxiety: 
One way to address range anxiety is to develop a strong FCS network to increase the chance of EVs to find charging stations conveniently, which requires significant investments. One the other hand, MCSs are reported in the literature as the key solution to reduce EV owners' range anxiety by providing additional chargers to them, which leads to less investments on FCS infrastructures \cite{noauthor_nation-e_2010}. 

\item[ii.] Charging availability: 
FCSs have a limited number of charging outlets and may not be able to cope with sudden high demands for charging \cite{chauhan2018scheduling}. 
Moreover, PuCSs include paid, discounted or free stations in publicly accessible parking garages and lots \cite{CLINTON2019102255}. According to Madison Gas and Electric, its public charging station users occupy the charging stations about 45 minutes longer on average than the time needed to charge their EVs \cite{doe2014evaluating}. In addition, EVs will likely be mainly charged at home in a garage \cite{bruninga2012overlooking} where, in many big cities where most of the households do not have their garages, charging may then take place outdoors \cite{6914759}.

By providing charging services at any time and location requested, according to the type of EVs, MCS can help to remove these obstacles.    

MCS could be also used for roadside assistance. With the increase in EV penetration, MCS can serve EVs out of charge on the road better and at a lower cost compared to tow service to the nearest charging station.
In other word, a MCS works according to the principle of a power bank for smartphones, but for EVs instead of \cite{noauthor_electrifying_nodate}.

\item[iii.] Charging time: 
 Generally, EV charging process is much slower than compared to the time required for ICEVs to fill up their tank at gas stations. The only charging technology which can compete is DCFC which can charge a given EV up to a 75\% state of charge (SOC) in around a half-hour \cite{noauthor_MYEV}. However, only 16\% of public charging stations so far are DCFCs \cite{noauthor_MYEV}. As a remedy, MCSs can help to fill this gap by providing faster charging options. Besides, if we consider the traveling time of going to a FCS, MCSs even save more time for a charging event by providing charging facilities at EV owners' locations \cite{cui2018multiple}.

\end{enumerate}

\subsubsection {Benefits to power grid} With increasing EV penetration, the number of charging requests will increase likewise. Depending on how fast the charging process is, the impacts on the power grid is different \cite{decker2012electric} and \cite{sun2016optimal}.
\begin{enumerate}
\item[i.] Slow charging: Most L2 charging stations are installed at workplaces, so it’s no surprise that L2 sessions peak when drivers plugin at two main times: when they arrive at work and when they come back after lunch \cite{noauthor_chargepoint3_2017}.
 FCSs can offer little charging flexibility. On the contrary, MCSs can store energy during off-peak hours and provide charging services for electric vehicles based on real-time charging demand \cite{wang2019location}.

\item[ii.] Fast charging: The high demand during fast charging can produce a significant voltage drop in the network and lead to system insecurity \cite{abdulaal2016solving}. Consequently, the power grid is prone to failure if several EVs are charged simultaneously; thus, it is essential to schedule EV charging processes properly \cite{sousa2018new,koufakis2016towards}. With increasing EV penetration and a higher number of DCFC connected to power grids, modifying the existing power grid infrastructure required significant investments \cite{decker2012electric} and \cite{mazidi2015optimal}. 
Therefore, a high number of FCSs is not possible to be provided with the current power grid.
By the expansion of MCSs, they can help the grid be exposed to a flatter demand profile and less need for investment to retrofit the infrastructure. In addition, employing MCSs fewer charging points to be installed at each location that suffers from FCS negative effect on the power grid.
\end{enumerate}
 
\subsubsection{Benefits to FCS investors} MCS can increase the utilization rate of FCS which can motivate investors to create new charging stations. 

\begin{figure}[htbp]
\centering
\includegraphics[width=0.9\linewidth]{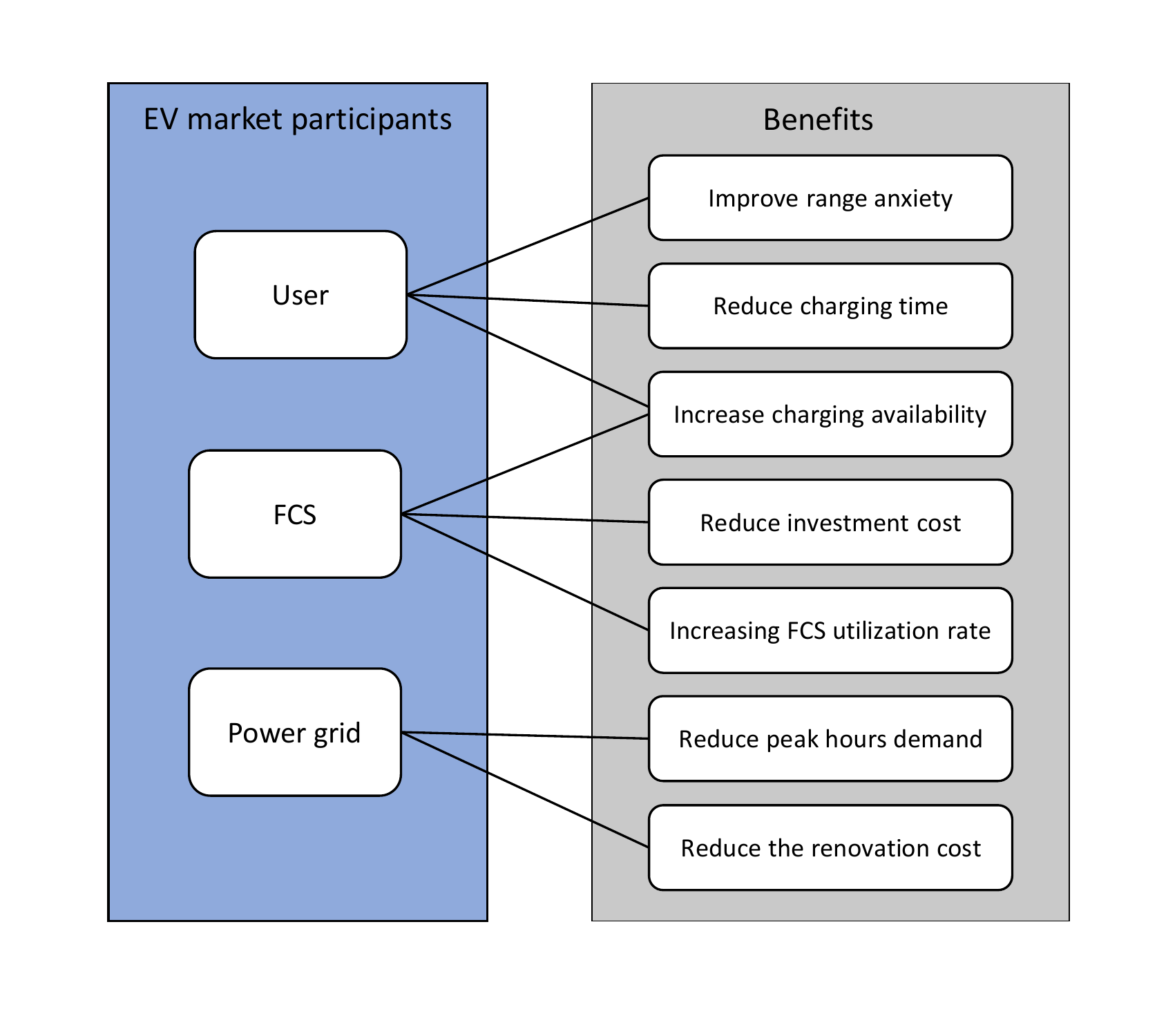}
\caption{MCS benefits from different perspectives}
\label{fig:benefits}
\end{figure}

\begin{enumerate}

\item[i.] Investment cost: It is difficult to determine whether to create a wide network of charging facilities to stimulate electric vehicles adoption or waiting for improving the rate of EV adoption before creating charging facilities  \cite{schroeder2012economics}.

As a solution, MCSs play a prominent role in addressing these needs. In a scenario to invest in FCS, using MCS, EVSE investors can find the best locations for FCSs before making significant investments to expand the charging network. Besides, it will be possible to set up many MCSs temporarily, exactly when and where they are required \cite{huang2014design}. This option could help the EVSE investors to have a better estimation of the number of charging requests and have more time to expand their fixed charging network. 

\item[ii.] FCSs' utilization rate: The low utilization rate of FCS, even in cities with a high density of EVs, is a serious obstacle slowing down the rate of return of investment in FCS \cite{WhitePaper}. In Shenzhen, China the number of charging piles has increased to 7962, though only 3697 charging piles can be used regularly, which is 46.3 percent of the total number \cite{Shenzhen}. Although MCSs lead to a decay in the penetration of FCSs, they increase the utilization of FCSs and their rate of return eventually. This will motivate more companies to invest in FCSs. 
\end{enumerate}
Fig.\ref{fig:benefits} depicts a summary on the benefits of MCS.

\subsection{Challenges of MCS}
Several significant engineering challenges must be overcome while designing MCS \cite{huang2014design}. For instance, designing the architecture of the mobile charger vehicle and the battery which can charge and discharge continuously considering the high cost of batteries and the battery life span \cite{atmaja2015energy}, the design challenge and cost related to the power electronic equipment \cite{badawy2015design} and the control system \cite{yu2009design} are the major concerns about MCSs. In terms of operation, balancing supply with demand to achieve optimal energy utilization of MCSs plus maximizing the power transfer efficiency are two main challenges with MCSs \cite{mou2019vehicle,wang2016spatio}. Designing of the charging navigation system for MCS to choose the stopping locations for providing charging services and avoiding extra trips to charge an EV or themselves is another challenge of MCSs \cite{li2019intelligent}.

In addition to these challenges, there are several other obstacles to operate MCS effectively, which are not adequately addressed in the literature. This paper elaborates these challenges as open research topics in the following sections.
\subsection{Planning}

EVs did not exist when the major part of the roads and grid infrastructures were built. To realize the shift from petrol and diesel to electricity as the main form of energy for transpiration, a reliable and accessible charging infrastructure is needed. This need has resulted in big investments in the EV charging realm \cite{noauthor_FreeWire2_2018}.
To ease the adoption of EVs, it is essential to mitigate the challenges and costs that come with creating a charging infrastructure \cite{noauthor_FreeWire_2018}. 
The cost of the battery and the cost of the carrier are the two major capital items of MCSs. However, the main challenge in MCS planning is the high cost of BESS.
In MCS planning and design, finding the optimal number of MCSs to cover charging requests has prominent importance. Optimizing the number of MCSs and the number of charging piles of each MCS to reduce the service delay has discussed in \cite{atmaja2015electric} and \cite{yang2012mobile}.  In \cite{atmaja2015electric} the results show that by increasing the number of charging piles form 1 to 4, the waiting time decreases from 89 minutes to 5 minutes.

\subsection{Operations/Scheduling/Dispatch}
An efficient procedure is proposed in \cite{atmaja2015electric}, \cite{chauhan2018scheduling}, and \cite{yang2012mobile} to increase the capacity of FCS temporarily by scheduling MCSs to serve additional EVs during peak hours. A dispatch algorithm is developed in \cite{atmaja2015electric}, where MCS with BESS and without BESS are dispatched to respond to requests from overloaded FCSs or directly from EVs' users.
The advantages of using MCSs in reducing waiting delay \cite{yang2012mobile,chen2018speed,li2010electric} and outage probability \cite{li2010electric} have been studied in the literature. 

In EV routing optimization, MCSs also make the EV routing problem much simpler. It transforms the EV routing problem with a high number of EVs looking for charging stations to an MCS routing problem with a lower number of MCS being dispatched to EVs. While EV routing has been addressed in several papers \cite{kedia2017review,shao2017electric}, a few papers have studied the operation of MCS.

\section{Mobile Charging Station In Practice}
Car manufacturers are investing heavily in EV innovation, hoping that EV demand will stay strong \cite{noauthor_chargepoint3_2017}. In addition to the big number of automakers and EV charging companies that provide charging facilities, several companies which are well known in other areas has a high investment in creating EVSE facilities. These companies including big oil and energy companies such as Shell and British Petroleum Company (BP) and engineering companies such as Siemens \cite{noauthor_20_electric}. However, some companies have gone further and offer new services include MCS which go to EVs to provide charging services.

\begin{figure*}[!htb]
  \centering
\begin{subfigure}{.4\textwidth}
  \centering
  \includegraphics[width=.9\linewidth]{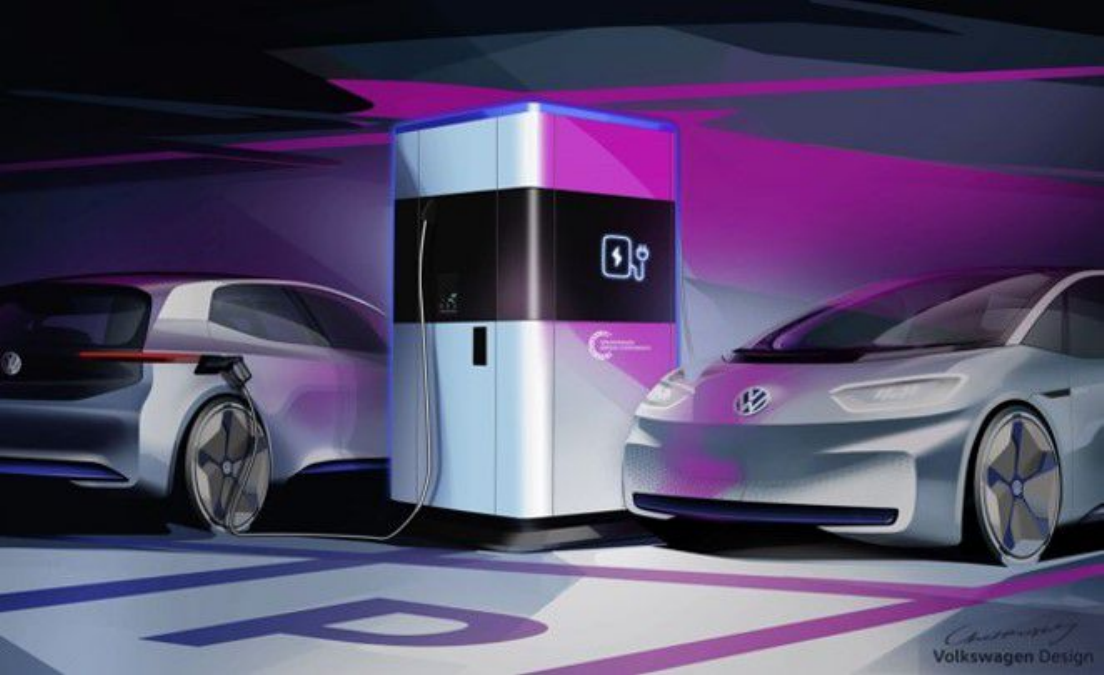}  
  \caption{}
  \label{fig:sub-first2}
\end{subfigure}
\begin{subfigure}{.4\textwidth}
  \centering
  \includegraphics[width=.9\linewidth]{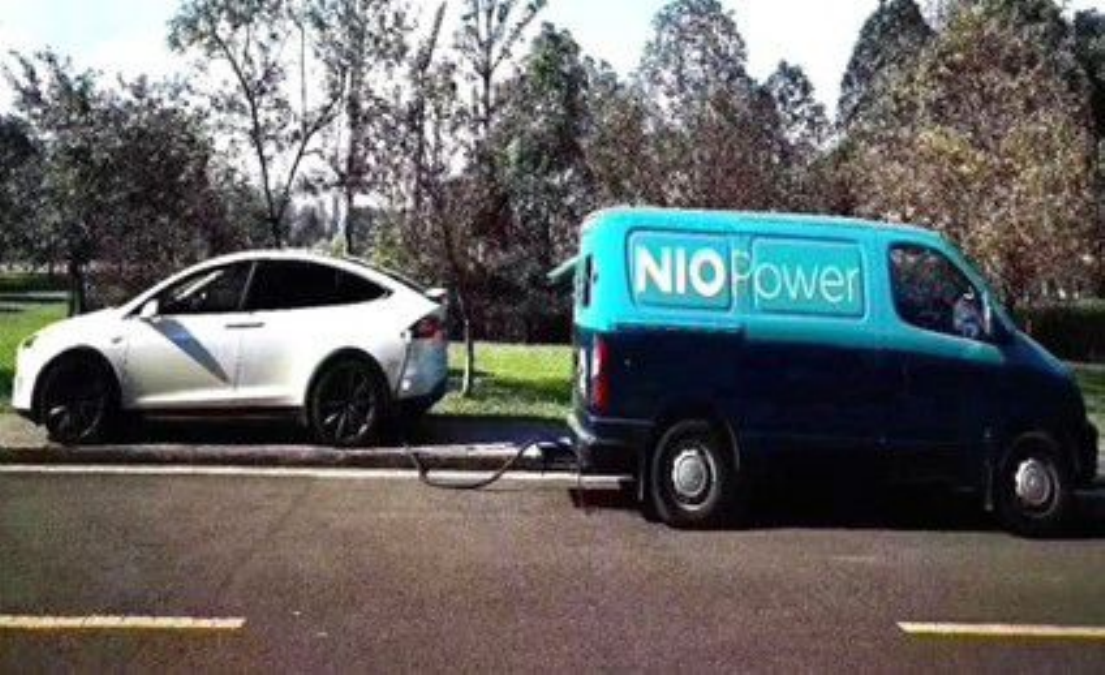}  
  \caption{}
  \label{fig:sub-second2}
\end{subfigure}
\caption{Mobile charging stations: (a) Volkswagen’s PMCS \cite{noauthor_electrifying_nodate} (b) NIO’s TMCS \cite{noauthor_nio_2018}}
\label{fig:PORTABLE}

\end{figure*}

{In 2010 Nation-E AG, the Swiss company for energy storage, introduced the first in the world, TMCS to help EVs to continue their path towards a nearby charging station \cite{noauthor_nation-e_2010}}. In terms of offering mobile charging services, Volkswagen and NIO are two car manufacturers considered MCS as a new charging service to charge electric cars \cite{noauthor_electrifying_nodate} and \cite{noauthor_nio_2018}. Recently, Tesla has also made its mobile super charging station \cite{teslamcs}. Moreover, MCSs can be offered by independent charging companies as the main form of charging or a service \cite{Andromeda}. These independent companies can be fixed charging services provider companies or just mobile service provider companies. 
In Fig. \ref{fig:PORTABLE} two model of MCSs has been shown.

\section{Open research topics}
While this paper has studied the feasibility, practicality, and benefits of MCSs, there are still many remaining issues that must be addressed. In the following we have identified these research topics.
\begin{enumerate}
\item[i.] From the perspective of distribution systems, the ability of MCS as energy storage could be helpful to reduce the negative influence of DCFC in the grid and reduce the total number of FCSs needed to cover charging requests in a specific location. As a result, considering the impacts of MCS grid instabilities and power quality needs to be studied. Moreover, it is necessary to consider the ability of charging methods such as MCSs in load management as possible controllable energy storage. In addition, the prominent role that MCSs could play in the EV market can be evaluated.

\item[ii.] It is also recommended to investigate different scenarios of operation of MCSs. To use MCSs more efficiently, it is necessary to consider a business model for them including the normal application MCS and their applications in the ancillary service market such as emergency purpose. { Besides, critical questions like the cost of a MCS compared to a FCS and round trip efficiency of MCS should be answered.}
Scrutinizing the impacts of EV owners’ strategies on the traffic conditions and pollution level considering MCS, optimal routing and charging of TMCS and the determination of stopping locations to operate MCS are the other available research topics.

\item[iii.] In the planning section, optimal charging station placement had always been a popular research topic of researches in EVs because it can have a considerable influence on range anxiety. However, by introducing MCSs there is a possibility to cut the costs of investment and increase the range anxiety by considering MCSs in charging placement. 
Moreover, designing the truck or portable charging stations in sections of power electronic, sizing and type of the battery and design of the truck could be other open research topics in the planning area. 
Besides, considering merging the new technologies of EV charging, the optimal design of a mobile battery replacement or designing possibility for a hybrid MCS that can support both mobile charging and battery swap technology are research topics, deserved to be discussed in the future.

\end{enumerate}

\section{Conclusion}
\label{Sec:Conclusion}
Mobile charging stations (MCS) in the state-of-the-art literature and in practice are studied. This paper also addressed various research aspects of MCS such as their benefits to EV owners and power grid, their challenges, and open research topics for future research. It concludes that further investigations are required in this research area. These research areas include, but are not limited to, optimal coordination between different charging methods including MCS, mitigation of adverse impacts of high EV penetrations on power grids through MCS, and impacts of MCS on market penetration of EVs. This paper serves as a step to understand the state-of-the-art in the area of MCS and as a foundations for new approaches for MCS to make EV charging as convenient and fast as filling up ICEV tanks with gas.

\bibliographystyle{IEEEtran}
{\footnotesize \bibliography{IEEEabrv,ref}}
\begin{IEEEbiography}[{\includegraphics[width=1in,height=1.25in,clip,keepaspectratio]{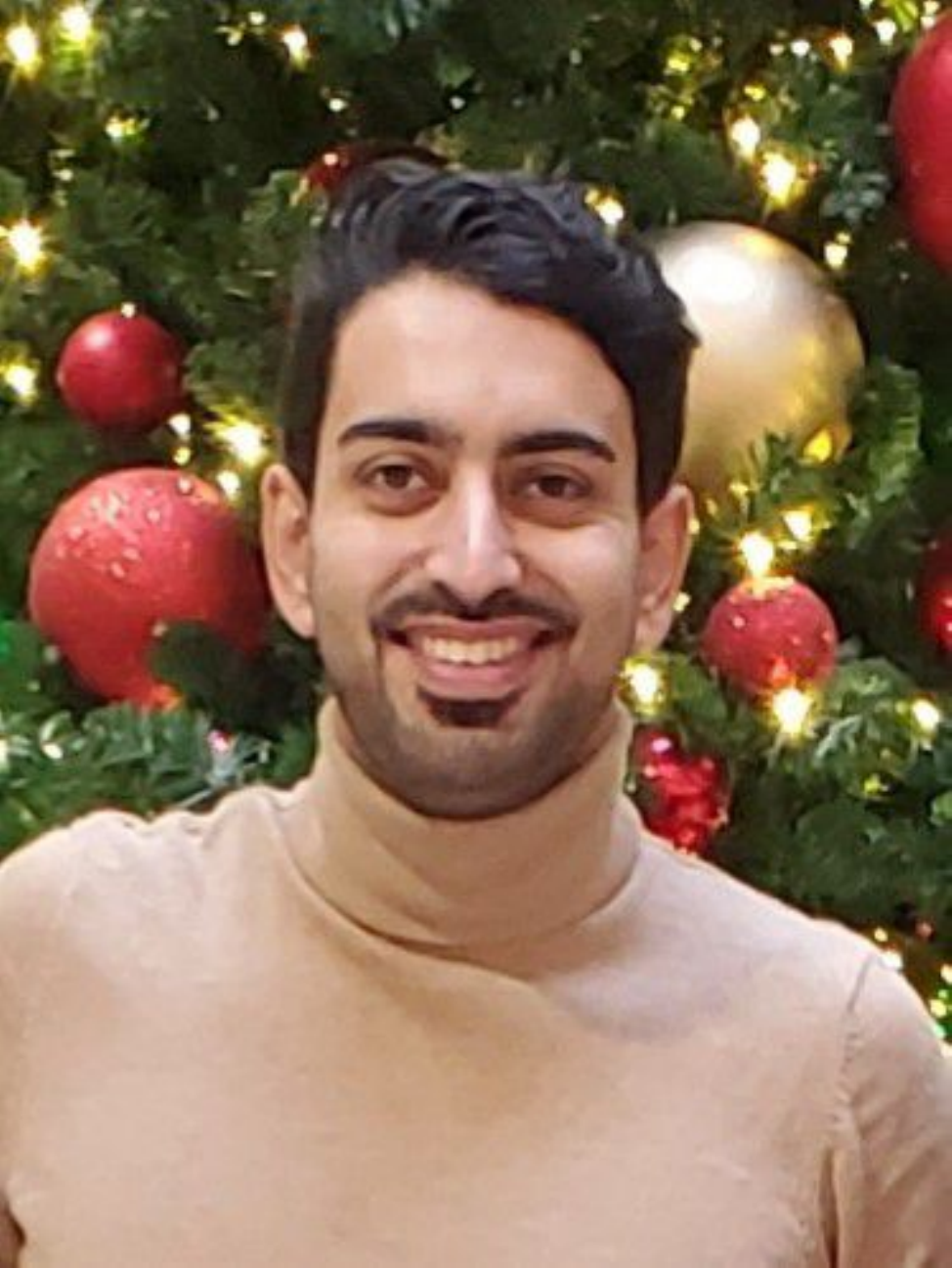}}]{Shahab Afshar}
 received the B.S. degree from the Islamic Azad University at Tehran, Iran, in 2014, and the M.Sc. degree from Tarbiat Modares University, Tehran, Iran, in 2016, both in  Electrical Engineering. He is currently working toward the Ph.D. degree at the University of Tennessee, Chattanooga, TN, USA. From 2017 to 2018, he was with the Power Systems Studies Research Department, Niroo research institute, Tehran, Iran. His research interests include transportation electrification, microgrid planning and operation, power system optimization.
\end{IEEEbiography}
\begin{IEEEbiography}[{\includegraphics[width=1in,height=1.25in,clip,keepaspectratio]{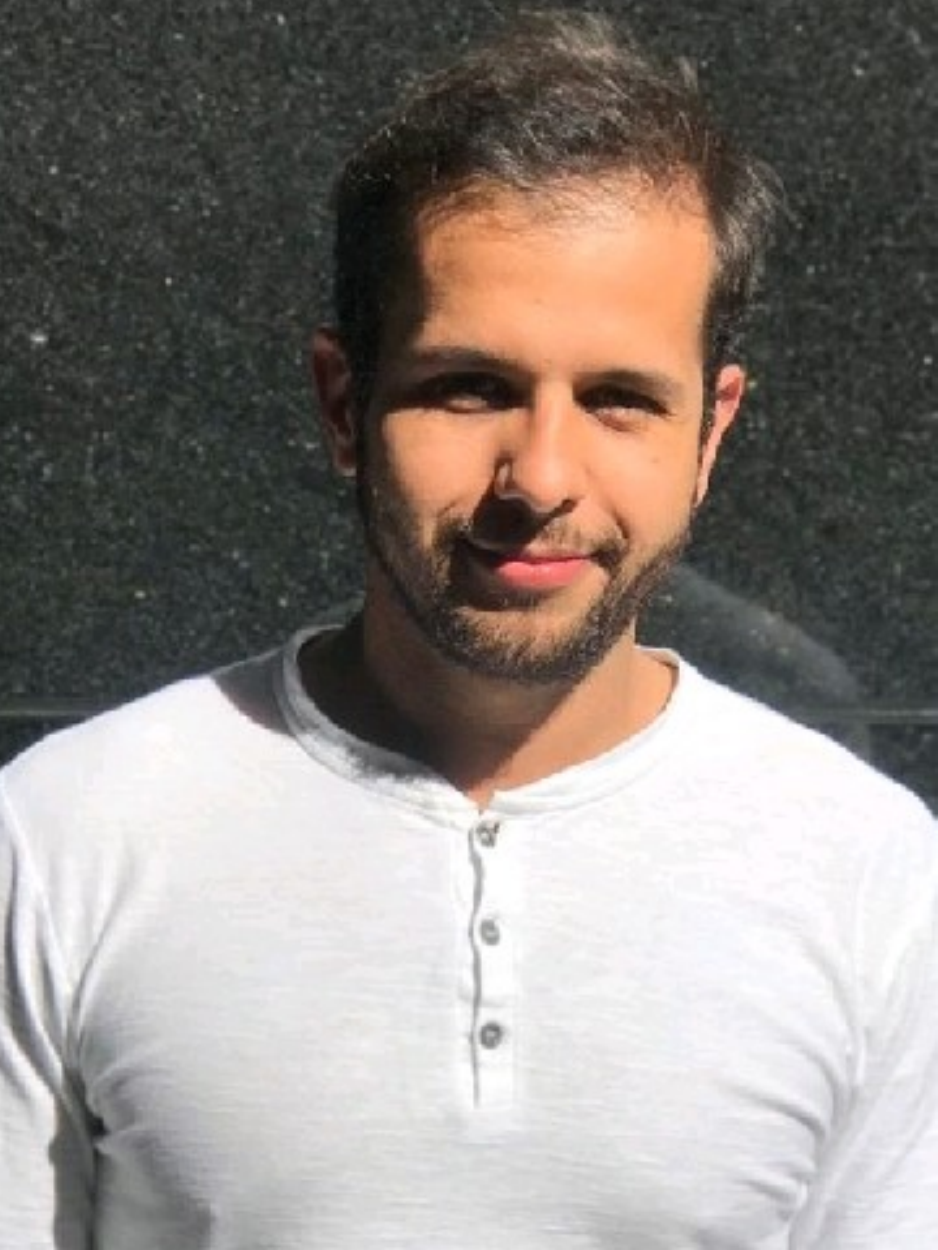}}]{Pablo Macedo}
 received the B.Eng. degree in electrical engineering from Federal University of Technology of Paraná, Brazil in 2018. From 2014 to 2015 he joined the University of California at San Diego (UCSD) as an exchange student and undergrad Assistant Researcher at the Center of Energy Research (CER). Currently he is pursuing his M.S. degree, in Electrical Engineering at University of Tennessee, Chattanooga, TN, USA. His research fields of interest is integration of distributed energy resources focused on frequency control design.
\end{IEEEbiography}
\begin{IEEEbiography}[{\includegraphics[width=1in,height=1.25in,clip,keepaspectratio]{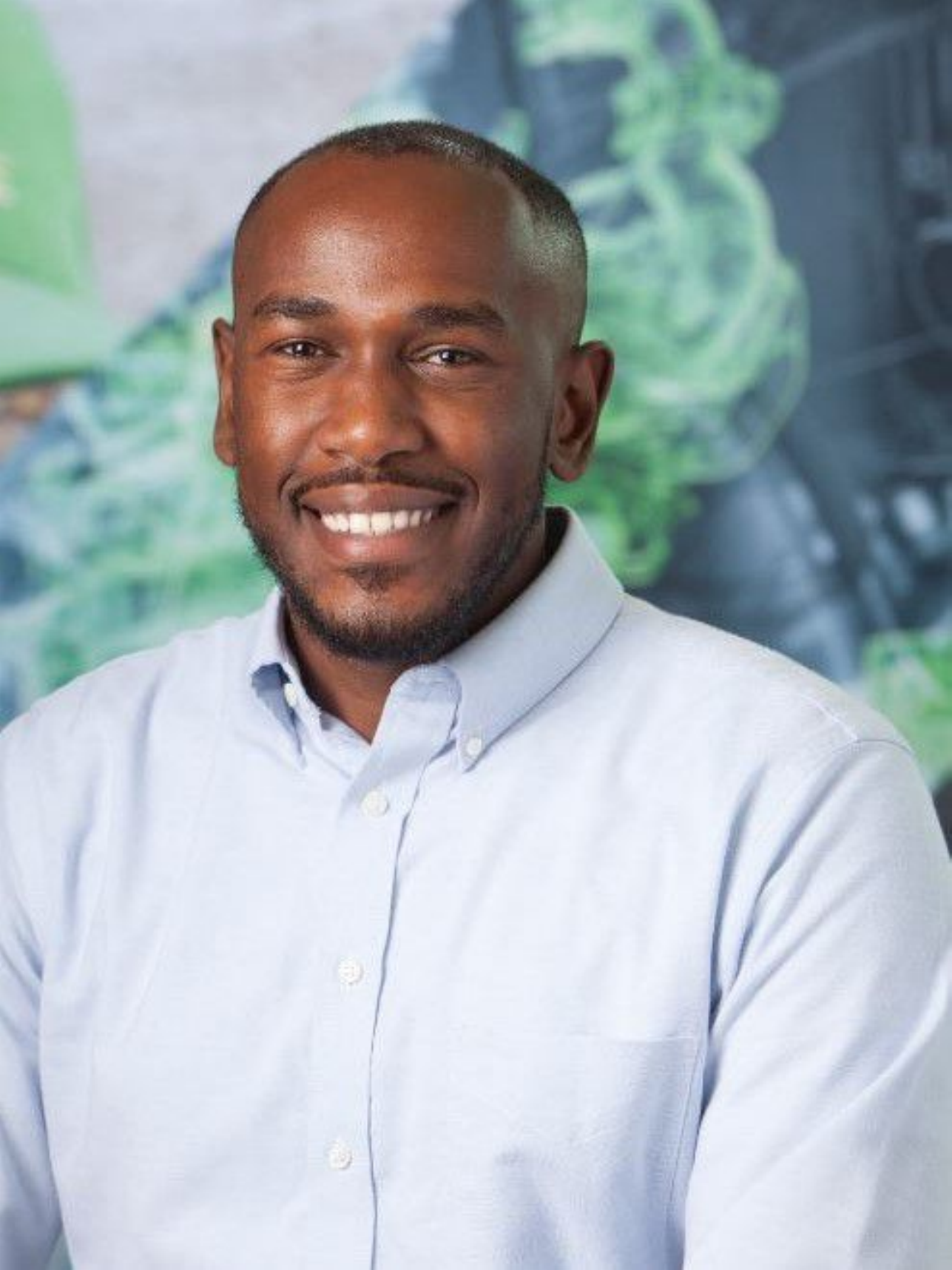}}]{Farog Mohamed}
received his B.S. degree in electrical and electronics engineering from the University of Khartoum, Sudan, in 2016 and he is currently working toward his M.S. degree in electrical engineering at the University of Tennessee, Chattanooga, TN, USA.  From 2016 to 2018, he joined the Sudanese Electricity Distribution Company as a research and development engineer. His research interests include grid integration of distributed energy resources, modeling of power electronics converters, energy storage systems, smart grid, and power system optimization.
\end{IEEEbiography}
\begin{IEEEbiography}[{\includegraphics[width=1in,height=1.25in,clip,keepaspectratio]{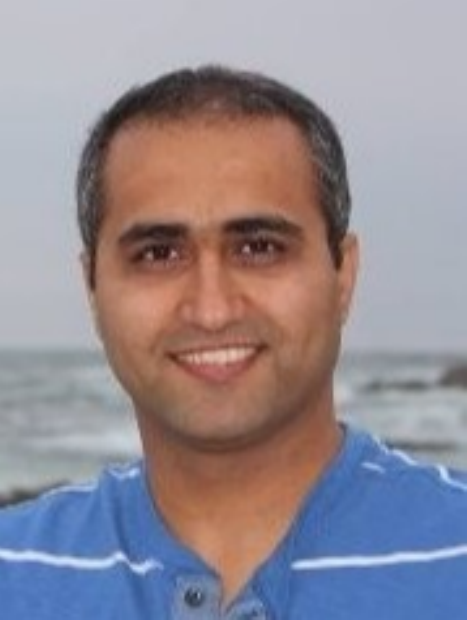}}]{Vahid Disfani}
received his Bachelor degree from Amirkabir University of Technology, Iran in 2006 and an M.S. degree from Sharif University of Technology, Iran in 2008, both in Electrical Engineering. He is currently an Assistant Professor and the Director of the ConnectSmart Laboratory, Department Electrical Engineering, University of Tennessee, Chattanooga. After earning his Ph.D. in electrical engineering from University of South Florida in 2015, he joined UC San Diego as a postdoctoral scholar. He has also contributed to several industry projects during his professional career at Iran Grid Management Company, his Ph.D. program and his two-year postdoctoral experience. His research  interests  include Power system optimization and control, Grid integration of renewable energy resources, Distribution system optimal voltage regulation, Power markets and power system economics.  
\end{IEEEbiography}

\end{document}